# Proposal for PhD topic:
# Two Neutron Correlation Study
# in Photofission of Actinides

Roman Shapovalov

Physics Department at
Idaho State Univerdity

Adviser: Prof. Dan Dale

April 1, 2012

**Abstract**

It is well known that two fission fragments (FF's) are emitted essentially back to back in the laboratory frame. That can be used widely in many applications as a unique signature of fissionable materials. However, such fission fragments are difficult to detect. The energy and angular distributions of neutrons, on the other hand, are easy to measure, and that distribution will carry information about the fission fragment's energy and angular spectra, as well as the neutron spectra in the fission fragment rest frame.

We propose to investigate the two neutron correlation yield resulting from two FF's as a function of different targets, the angle between the two neutrons and the neutron energies. The preliminary calculation of the two neutron correlation shows a huge asymmetry effect: many more neutrons are emitted anti-parallel to each other than parallel to each other. That asymmetry becomes even more if the energy cut on each neutron is done. This study will potentially permit a new technique for actinide detection for homeland security and safeguards applications as well as improve our knowledge of correlated neutron emission.

# Contents

# Chapter 1

# Statement of the physics problems

## 1.1 Simple summary of fission physics

The physics of photofission is well described in many books [1, 2]. The overall process can be schematically represented as shown in Fig. 1.1. What we are going to discuss here is up to the time scale of about $10^{-14}$ to $10^{-13}$ sec when the prompt neutrons are emitted from the fully accelerated fragments and completely ignore all the following processes where the prompt gammas and delayed $\beta$, $\gamma$ and $n$ are emitted. We will only touch on some specific information we will need to understand the underlying physics in the proposed two neutron correlational study. That mechanism is, of course, in some sense, an approximation, because we do not count possible "scission" neutrons emitted at the instant of fission [10]. What we assume here is that all neutrons are emitted from fully accelerated fission fragments.

It has long been known that the photofission reaction with a heavy nucleus in the energy range of the giant dipole resonance goes through the intermediate compound nucleus. That intermediate nucleus is in an excited state followed by the emission of two fission fragments:

$$\gamma + A \to A^* \to FF_1 + FF_2 + TKE \tag{1.1}$$

where TKE is the total kinetic energy which will be shared by the two fission fragments. In general, the TKE will be a function of the fragment mass which has been measured by several authors [19, 20, 21, 22] as seen in Fig. 1.2 [17]. Because the fission fragments are essentially non relativistic, the TKE will be distributed proportional to their mass ratio as:

$$\frac{T_1}{T_2} = \frac{M_2}{M_1} \tag{1.2}$$

where $T_1$, $T_2$ are the kinetic energies of fragments 1 and 2 such that $TKE = T_1 + T_2$ and $M_1$, $M_2$ are their rest masses correspondingly.

The typical mass distribution at the energy range not too far from the threshold barrier is shown in Fig. 1.3 [14]. It is symmetric about $A = 120$ and for every heavy fragments there is a corresponding light one, but the fission with two equal mass fragments is less probable by a factor of about 200. It is interesting, that as the energy of incident $\gamma$'s increases, the masses of two FF's tend to be more equal [15].

The angular distribution of individual FF's can be explained according to A. Bohr's fission channel concept [5] and briefly described by R. Ratzek et al. [11] with regards to photoinduced
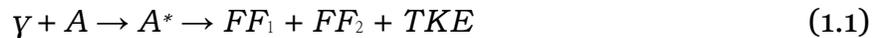

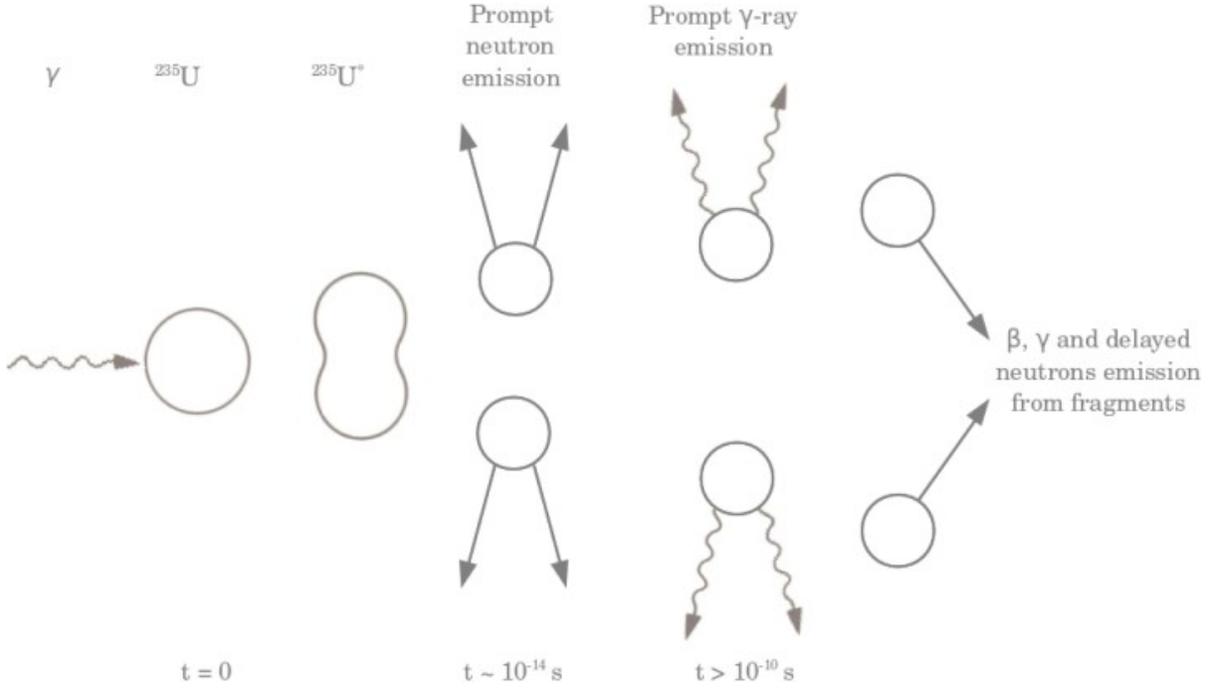

Figure 1.1: Schematic representation of the fission process in uranium. Neutrons are emitted from fully accelerated fission fragments. The time scale gives the orders of magnitude only [26].

reactions. If we restrict ourselves to the photofission of an even-even nucleus ($J^\pi = 0^+$, such as $^{238}U$) and consider only electric dipole (E1) transitions, the angular distribution of fission fragments can be written as [11]:

$$W(\Theta) = A_0 + A_2 P_2(cos\Theta) \qquad (1.3)$$

The angular distribution coefficients $A_0$ and $A_2$ depend on the transition state (J,K), where K is the projection of the total spin J on the symmetry axis of the deformed nucleus. For $J = 1$, $K = 0$, we have $A_0 = \frac{1}{2}$, $A_2 = -\frac{1}{2}$ and for $J = 1$, $K = 1$, we have $A_0 = \frac{1}{2}$, $A_2 = \frac{1}{4}$. $P_2(cos\Theta) = \frac{1}{2}(2 - 3\sin^2\Theta)$ is the Legendre polynomial. Qualitatively, the angular distribution of the fission fragments can be explained if we consider the nuclear excitation as a collective motion of neutrons against the protons [4]. Because the incident gammas are a transverse wave, that will cause protons to oscillate against the neutrons in the direction of electric field **E** followed by the splitting of nucleus into the two fission fragments.

Some simple considerations of kinematic of reaction 1.1 can clarify some important moments. In the first step the incident gammas interact with heavy nucleus A resulting in compound intermediate state A*. For such a step, if the energy of incident gammas is small, say below or about 20 MeV, after applying the momentum conservation law, we can easily see that the excited nucleus A* is almost in rest. Because of that, and applying the momentum conservation law to the last step of the reaction, we conclude that the two FF's are flying away almost in opposite direction as seen in the laboratory frame. This simple conclusion is very important and can be used widely in many applications as a unique signature of fissionable materials.

After about $10^{-14} - 10^{-13}$ sec the fission fragments will emit neutrons. As was already mentioned, it was assumed that all prompt neutrons are emitted from fully accelerated fragments, and there are no so called "scission" neutrons emitted at the time of fission. The important parameter to be considered here is the total excitation energy (TXE) of the intermediate nucleus $A^*$. That total excitation energy will be shared among light and heavy fragments, and the exact form of such distribution is the open question. However, there is strong evidence [23, 24] that when the excitation energy is relatively low the light fragments will acquire the larger part of that shared energy. Those excited fission fragments can release energy and angular momentum by emitting prompt neutrons and prompt γ rays as well, but it can be assumed that the initial energy release is completely due to the neutron emission [17]. Because the excitation energy of the fission fragments is large in comparison with the lowest lying nuclear levels, the statistical model to analyze the neutron emission spectrum can be applied [3]. Using that approach, to a very good approximation, the angular distribution of prompt neutrons is isotropic in the center-of-mass frame of the fission fragments. The energy of the evaporated neutrons can be described by the Maxwell distribution with the spectrum temperature $T$:

$$\rho(E_n) = \frac{E_n}{E_n} \exp\left(-\frac{E_n}{T}\right) \quad (1.4)$$

where $E_n$ is the neutron kinetic energy in the center-of-mass fragment frame.

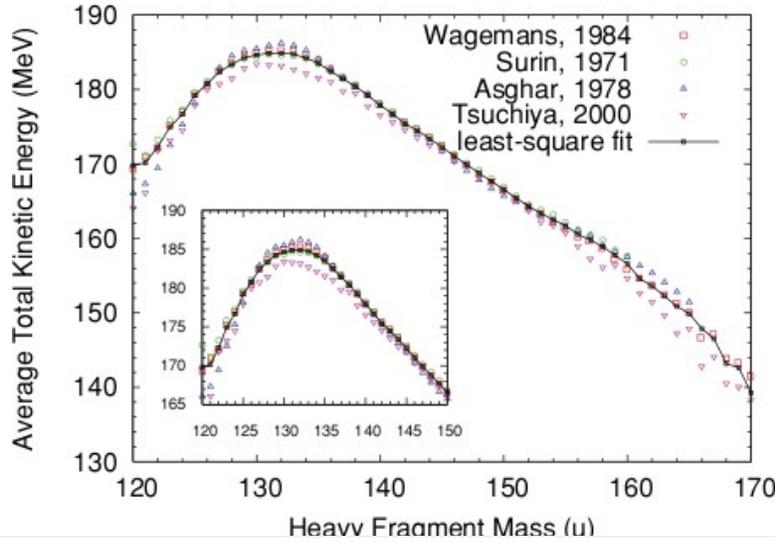

Figure 1.2: The average TKE as a function of the heavy fragment mass. The solid line is the result of a least-square fitting of the experimental data sets [17].

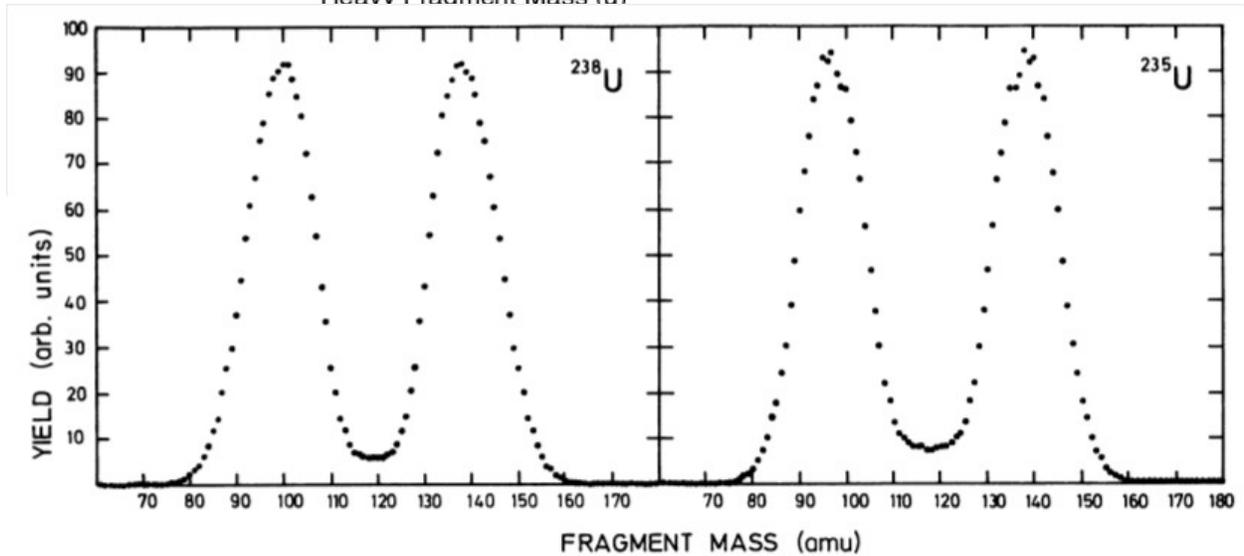

Figure 1.3: Integrated fission fragments yield (arbitrary units) versus fragment mass for the photofission of $^{238}U$ and $^{235}U$ with 25-MeV bremsstrahlung [14].

After the first neutron is emitted the second one will be emitted and so on until the excitation energy of the fragments becomes less then the neutron separation energy. Finally, the rest of the excitation energy can be released by prompt $\gamma$ ray emission. However, what we assume here is that only one neutron is emitted from each of the fully accelerated fission fragments.

Below is a short summary of the photofission reaction mechanisms discussed above which will be used in the following section to discuss the idea of the proposed two neutron correlation:

- two fission fragments recoil essentially back to back.

- the angular distribution of the prompt neutrons is isotropic in the center-of-mass frame of the fission fragments with a statistical energy distribution.

- each fully accelerated FF emits only one neutron.

## 1.2 Idea of 2n correlations

Let's start to count how many FF's pairs are going antiparallel and how many FF's pairs are going parallel to each other. Because two fission fragments recoil back to back, the FF's asymmetry would be, of course, infinity (there are no two FF's going parallel to each other):

$$A_{FF} = \frac{FF\text{'s antiparallel}}{FF\text{'s parallel}} = \infty \quad (1.5)$$

where *FF's antiparallel* is the number of FF's pairs going in antiparallel direction and *FF's parallel* is the number of FF's pairs going in parallel direction.

The problem here is that fission fragments are very difficult to detect. For a target thicker than a few $mg/cm^2$, due to their heavy ionization loss, almost all fission fragments will stop inside the target. On the other hand the neutrons emitted by these fission fragments will fly outside of the target and could be easily detected. The question here is whether or not the angular asymmetry of fission fragments (they are always back to back) is manifest in the angular distribution of prompt neutrons. In order to answer this, we propose to measure the two neutron angular and energy distributions with the ultimate goal of calculation the two neutron asymmetry:

$$A_{2n} = \frac{2n\text{'s antiparallel}}{2n\text{'s parallel}} \quad (1.6)$$

where *2n's antiparallel* is the number of 2n's pairs going in antiparallel direction and *2n's parallel* is the number of 2n's pairs going in parallel direction as seen in the LAB frame.

If we take a typical 1 MeV neutron in the center-of-mass frame of the fission fragment it will travel with the speed of about 4.6% of the speed of light. The angular distribution of neutrons in this frame will be essentially isotropic as was discussed previously. If we take two fission fragments with typical mass numbers $A_1$ = 95 and $A_2$ = 143 they will travel with the speed of about 4.6% and 3.0% of the speed of light correspondingly, and they will fly away in the opposite direction. The energy and angular distribution of neutrons observed in a LAB frame will be a superposition of these two spectra: 1) the spectrum of neutrons in the fission fragment rest frame and 2) the spectrum of the fission fragments.

The expected 2n correlation asymmetry could be thought of as a product of asymmetry of two fission fragments $A_{FF}$ (eqn. 1.5) times a washing effect due to isotropic angular distribution of neutrons in the fission fragment rest frame $W_n$ times a washing effect due to neutron multiple scattering effect inside the target and surrounding materials $W_{scat}$:

$$A_{2n} = A_{FF} \cdot W_n \cdot W_{scat} \qquad (1.7)$$

Because the first factor is a large we can expect that the total two neutrons asymmetry as measured in laboratory frame (eqn 1.6) would be the sufficient to observe.

# Chapter 2

# Brief review of what has been done

The first ever measurements of photofission fragment angular distributions were performed on Thorium in 1952 - 1954 by several authors [6, 7, 8] and were summarized and briefly discussed by Winhold and Halpern in 1956 [9]. It was found that the observed angular distribution has the form $a + b\sin^2\Theta$ (Fig.2.1) and the ratio $b/a$ depends on the energy of the photons producing the fission, on the particular fissionable target being irradiated, and on the particular fission fragments being observed. It was found that the photons in the giant resonance region produce essentially isotropic fission and the anisotropic fission is due solely to photons within about 3 MeV of the fission threshold. As can been seen from Fig 2.2 the anisotropies in $Th^{232}$ and $U^{238}$ decrease rapidly with increasing electron energy and there are not any anisotropies for $U^{235}$ was measured. That was discussed and analyzed using the Bohr model of collective motion [5].

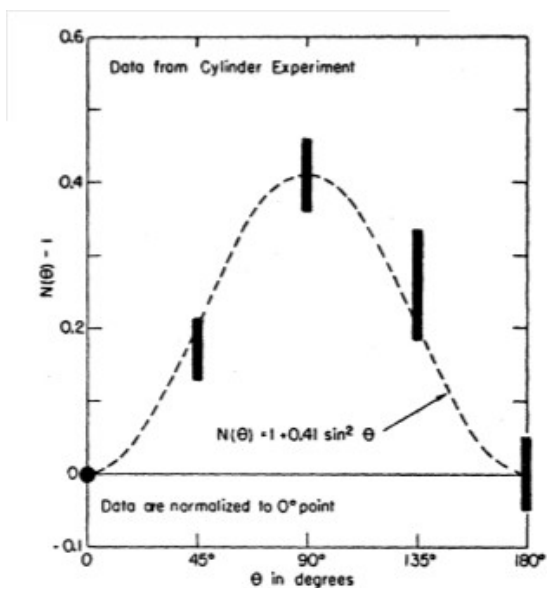

Figure 2.1: The angular distribution, $N(\Theta)$, of fission fragments from $Th^{232}$ caught at the angles $\Theta$ to the x-ray beam. The x-ray beam was produced in a thick lead target by an electron beam whose spectrum was centered at 13 MeV and was about 5 MeV wide [9].

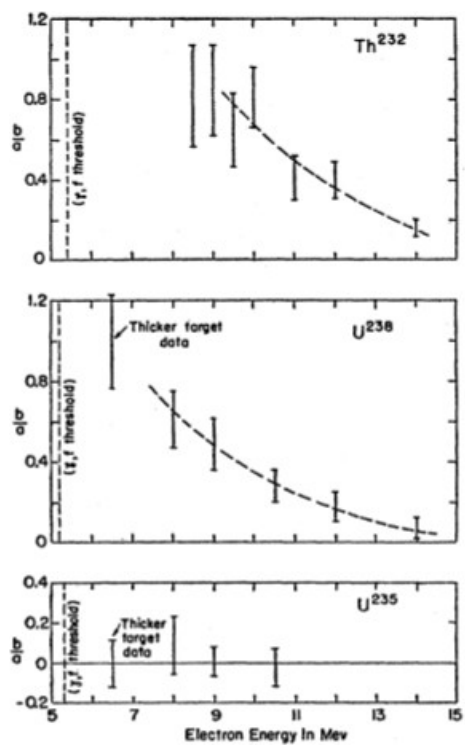

Figure 2.2: The anisotropy in the photofission of three targets. The angular distributions were all assumed to be the form $a + b\sin^2\Theta$ [9].

Years later in 1962 the neutron angular and energy distributions were measured by Bowman et all [10]. They analyzed the spontaneous fission of $^{252}Cf$ by using the time of flight technique to measure the neutron angular and energy distributions in coincident with the fission fragments. The experimental data were analyzed under the assumption that there are no 'scission' neutrons and there are 10% of 'scission' neutrons. The last assumption in general gives better agreement with the measured data as can been seen from the Fig. 2.3. The calculated energy spectrum of neutrons in the CM frame is presented in Fig 2.4. The large dots represent the neutrons emitted in the direction of the light fragments and the triangles represent the neutrons emitted in the direction of the heavy fragments. The smaller dots were obtained from measured neutrons emitted in the backward direction from the light fragments. The curve for light fragments was reduced by the factor 1.16, which is the ratio of the number of neutrons from the light fragments to the number from the heavy fragments. The results can be explained well by assumption of isotropic evaporation of neutrons from the fully accelerated fragments.

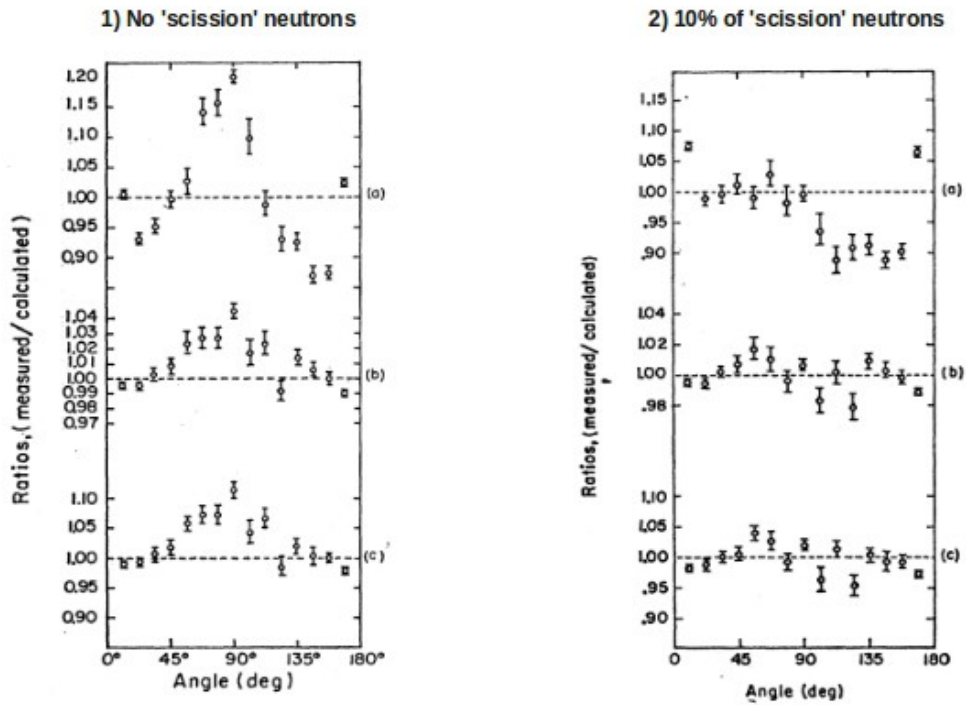

Figure 2.3: The ratio of measured to calculated values for (a) numbers of neutrons (b) average velocities, and (c) average energies as a function of angles.

Further measurements of angular and energy distributions of fission fragments and neutrons from the spontaneous fission of $^{252}Cf$ were made by Budtz-Jorgensen and Knitter in 1988 [12]. The measured neutron energy spectrum (Fig. 2.5) is in very good agreement with the Maxwell distribution in the energy range below 20 MeV energy point with the temperature parameter of $T = 1.41 \pm 0.03$ MeV. The neutron angular distribution recalculated in the fission fragment rest frame integrated over all neutron energies and normalized to unity is plotted in Fig. 2.6. The results confirm the isotropic neutrons' angular distribution suggested by many authors in most modern theoretical models. The obtained angular anisotropies are compared by authors with data obtained by Bowman *et al.* [10] as a function of fission neutron energy and is presented in Fig. 2.7. There is good agreement between both measurements up to about 4 MeV and significant discrepancy above that point. The solid line is a theoretical line calculated with the assumption that there are no 'scission' neutrons and is in good agreement with the Budtz-Jorgensen measurements.

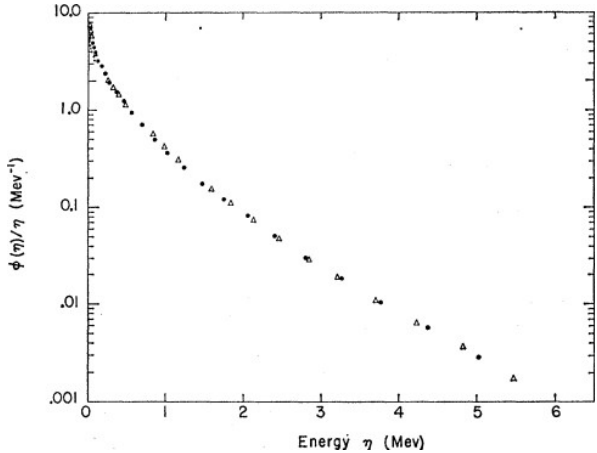
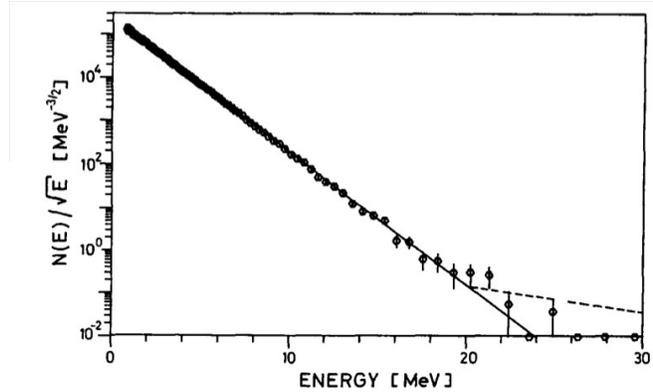

Figure 2.4: The center-of-mass neutron energy spectrum $\varphi(\eta)$ (CM) divided by $\eta$.

Figure 2.5: Fission neutron energy spectrum divided by the square root of the neutron energy versus the neutron energy. The solid line is Maxwell energy distribution.

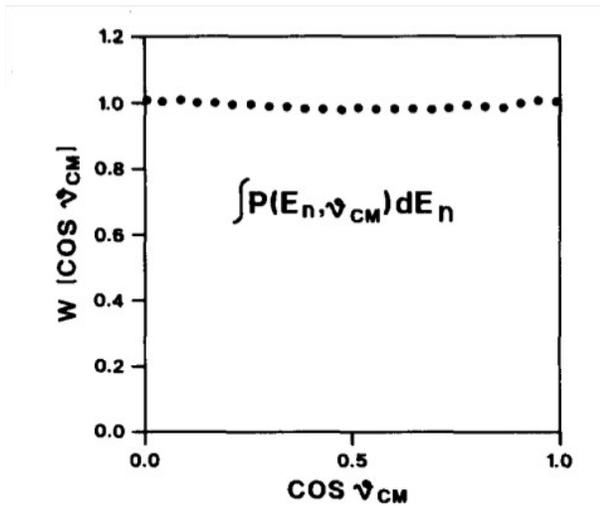
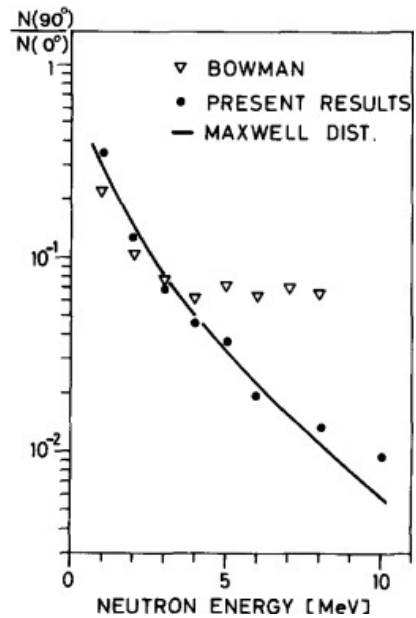

Figure 2.6: Fission neutron angular distribution in the fragment center-of-mass system integrated over all neutron energies

Figure 2.7: Fission neutron intensity ratio $N(90^o)/N(0^o)$ is plotted versus the fission neutron energy.

# Chapter 3

# Our experimental set-up

We plan to use the HRRL LINAC to construct the beamline to produce the bremsstrahlung photons. From private communication [27], that machine can supply the 20 ns and higher pulse width with about 10-80 mA peak current. That will give to us the freedom to adjust the beam parameters to satisfy the desired condition to have the one fission per pulse as will be described in the following section. Because such a low rate is needed, the main advantage of HRRL LINAC is, of course, the high repetition beam pulse rate of 1000 Hz that will permit an increase the statistics as compared with the other machines available in IAC.

The production of unpolarized photons is a well known technique and is widely described in the literature [13]. When electrons strike the radiator, that results in the bremsstrahlung radiation in the forward with respect to the beam direction. The typical energy spectrum of bremsstrahlung photons for the 7 MeV endpoint energy is shown in Fig 4.6.

Such a beam of unpolarized photons will be used to measure the two neutron correlation yield as a function of different targets, the angle between the two neutrons and the neutron energy. The time of flight (TOF) technique will be used to identify neutrons and to measure their energy, with the start signal coming from the accelerator beam pulse. Fig. 3.1 shows a typical time of flight spectrum from photodisintegration of the deuteron measured from previous HRRL runs.

A typical 1 MeV neutron travels about 5% of the speed of flight. If we take the neutron

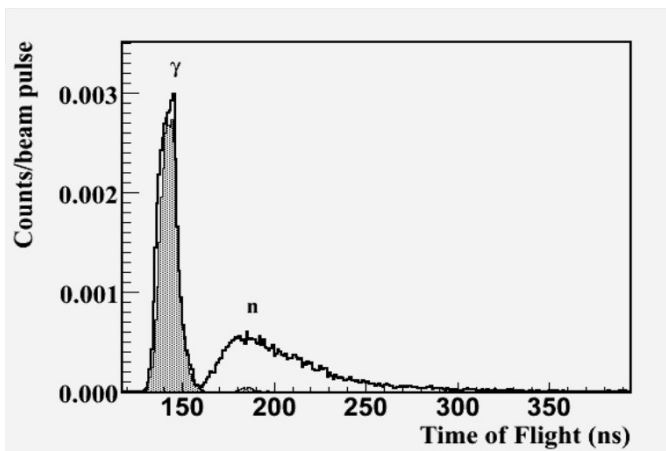

Figure 3.1: Typical TOF spectrum from photodisintegration of deuteron measured from previous HRRL runs. The distance from target to detector is about 2 m. The spectrum illustrate the ability to distinguish gammas peak from neutrons one.

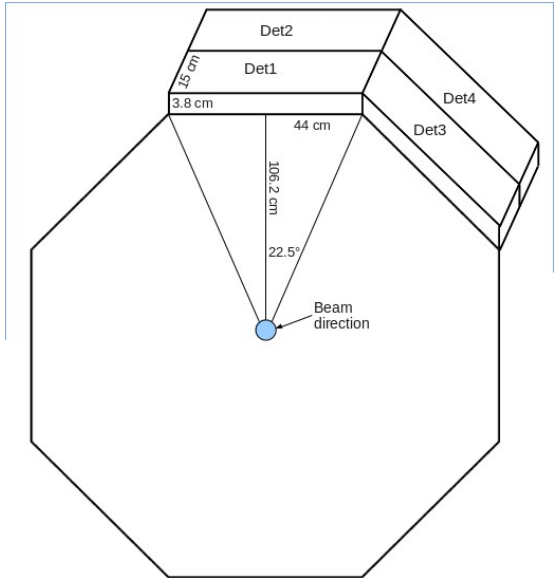

Figure 3.2: Possible detector geometry to measure the two neutron correlation yield. Total 16 neutron detectors are placed at the angle of 90 degree with respect to the beam. The detector size is 15 cm × 88 cm × 3.8 cm.

detector located 1 m away from the target, that will correspond to the TOF equal to:

$$\frac{1\ m}{0.05 \times 3 \cdot 10^8\ m/s} \approx 67\ ns$$

The TOF of gammas scattered from the target and flying with the speed of light $c$ will be around 3.3 ns. That will allow us to distinguish neutrons from gammas. By converting the measured time of flight of neutrons to their velocity we will be able to reconstruct the neutron energy. Of course, the error in neutron energy will depend on the LINAC pulse width. For HRRL the minimum pulse width, as was mentioned above, is about 20 ns and that will limit the precision with which we will be able to measure the neutron energy. To reduce such a kind of error the distance from target to detector could be increased up to about 2 m.

Because the one fission per pulse condition is required, the neutron detectors with the large area are needed. We currently have 16 plastic scintillators with the size of about 15 cm ×88 cm ×3.8 cm that corresponds to an area of about 15 cm ×88 cm = 0.132 m$^2$. As will be shown later for the uranium-238 target, the neutrons are emitted mostly perpendicular to the beam direction (Fig. 4.2). To maximize the 2n correlation yield such plastic scintillators will be placed at the angle of 90 degree with respect to the beam surrounding the target. Further thinking and calculation about the detector location should be done but, in principal, that will allow almost $2\pi$ cover as can be seen from Fig 3.2. Two PMT's will be symmetrically attached to each end of each detector. To increase the collected light from the detector especially at the area close to the ends, the non-scintillating plastic transparent to the visible and UV light will be placed between the detector and PMT.

Assume the neutron hits the detector at some distance $y$ from the first PMT as shown in Fig 3.3. Two techniques to find the position $y$ can be used here. The first method is as follows. The amplitudes $A_1$ and $A_2$ detected by PMT$_1$ and PMT$_2$ correspondingly will be proportional to the distances $y$ and $(l - y)$ that light travels as follows:

$$A_1 = I_0 e^{-ay}$$

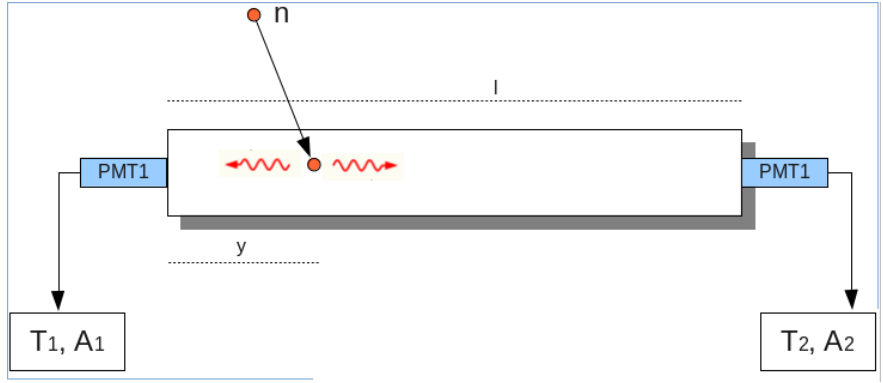

Figure 3.3: Neutron detector with two PMT's attached to both each end. Neutron n hits the detector at distance y from first PMT. The amplitude signals $A_1$, $A_2$ and TOF signals $T_1$, $T_2$ are measured from $PMT_1$ and $PMT_2$ correspondingly.

$$A_2 = I_0 e^{-a(l-y)}$$

where $l$ is the detector length and $a$ is the attenuation constant. If we take the natural logarithm of the ratio of $A_1$ and $A_2$, the distance $y$ where the neutron hit the detector becomes:

$$y = \frac{l}{2} - \frac{1}{2a} \ln \frac{A_1}{A_2} \qquad (3.1)$$

The other method we can use here is the timing technique. The TOF $T_1$ and $T_2$ detected by $PMT_1$ and $PMT_2$ correspondingly can be calculated as follows:

$$T_1 = \frac{L}{c} + \frac{yn}{c}$$

$$T_2 = \frac{L}{c} + \frac{(l-y)n}{c}$$

where $l$ is as before the detector length, L is a distance the neutron travels from the target to the detector, $c$ is the speed of light and $n$ is the index of reflection of scintillator material used in the detector. Taking the difference of $T_1$ and $T_2$ the position $y$ can be found easily:

$$y = \frac{c}{2n}(T_1 - T_2) + \frac{l}{2} \qquad (3.2)$$

Both techniques can be used to calculate the position where the neutron hits the detector. However the last method looks more simple and preferable in the following sense. In the first method the amplitudes of both PMT's for each detector should be measured. So, we need to know the relative efficiency of both PMT's to be able to match the gain of the signals. To find the energy of a neutron, in addition, the TOF spectrum measurements, are needed as well. The two independent channels of an acquisition system are needed in that case for each PMT's. In the last timing technique method the only TOF measurements for each PMT are required. That will allow to find the position $y$ as described by the formula 3.2 as well as the neutron energy by converting the TOF to the neutron velocity. So only one acquisition system channel will be needed in the last case.

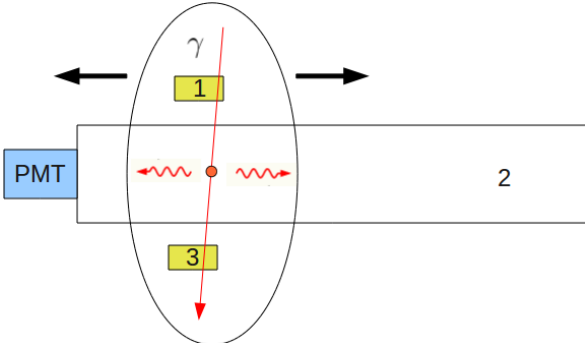

Figure 3.4: TOF measurements set-up. The triple coincidence between detectors 1, 2 and 3 from the cosmic ray was used as a start signal to measure the time as a function of distance.

Some preliminary TOF measurements with 1 PMT attached to the end of the detector were performed and the results presented in Fig. 3.5. Two small plastic detectors 1 and 2 were placed above and under the "big" plastic detector and were moved along the "big" one (Fig. 3.4). The triple coincidence between detectors 1, 2 and 3 from the cosmic ray was used as a start signal to measure the time as a function of distance. The results show the ability to identify the source position as a function of measured TOF. The calculated average speed of light inside the scintillator is about 7 cm/nsec that is about 4 times less than the speed of light. Also note the minimum distance from the source to PMT where the data were collected is about 15 cm. Below that point no signal was detected. That is simply because as was found later, there is no scintillating materials in this region.

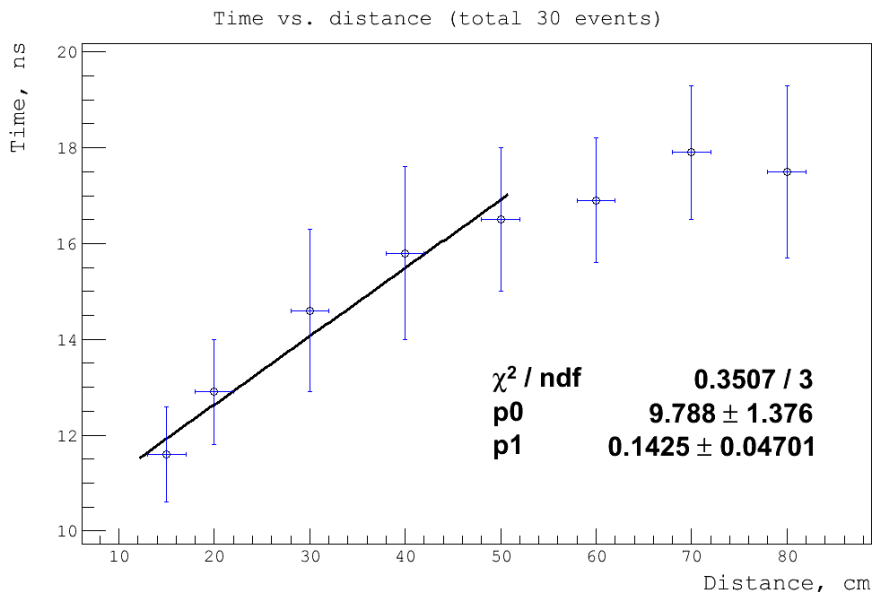

Figure 3.5: TOF measurements with 1 PMT attached to the end of detector.

# Chapter 4

# Expected results

## 4.1 Asymmetry calculation

To estimate the expected asymmetry in 2n correlations, a Monte-Carlo simulation was performed. A total number of 10 million fission events was simulated. Each neutron was sampled up to 10 MeV in the fission fragment rest frame. The following was assumed here:

- The uranium-238 with $J = 1$ and $K = 0$ is used as the fissionable target.

- The incident gammas are an unpolarized wave.

- The fission fragment mass distribution is sampled uniformly between $85 < A < 105$ and $130 < A < 150$

- A fixed amount of total kinetic energy of 165 MeV is given to the two fission fragments and is distributed between them proportional to their mass ratio

- Each fission fragments emit one neutron. There are total two neutrons, marked as $a$ and $b$ for each fission event. Neutrons are emitted isotropically in the center-of-mass of fully accelerated FF's with the energy distribution given by:

$$N(E) = \sqrt{E} \exp\left(-\frac{E}{0.75}\right) \tag{4.1}$$

This reproduces the laboratory neutron energy distribution as measured with (n,f) channel.

- Two recoiled fission fragment are emitted back to back. The fission fragment angular distribution is sampled according to:

$$W(\Theta) = \frac{1}{2} - \frac{1}{2}\left(\frac{1}{2}(2 - 3\sin^2\Theta)\right) = \frac{3}{4}\sin^2\Theta \tag{4.2}$$

for $J = 1$, $K = 0$.

After both angular and energy distributions of neutrons and FF's were sampled using the assumptions above, neutrons were boosted from the fission fragments rest frame into the laboratory frame. The energy and direction of neutrons a and b for every fission event were recorded in the LAB frame.

To be sure that the simulated algorithm is correct, some preliminary results of the above described simulations are discussed below.

The energy spectrum of the sum of the kinetic energies of the two neutrons *a* and *b* emitted by fully accelerated fission fragments as seen in laboratory frame is plotted in Fig. 4.1. Because the typical neutron energy in the fission fragment rest frame is about 1 MeV and the spectrum above is the spectrum of the sum of two neutron energies, the peak value at about 2.4 MeV looks reasonable after the boost into the LAB frame.

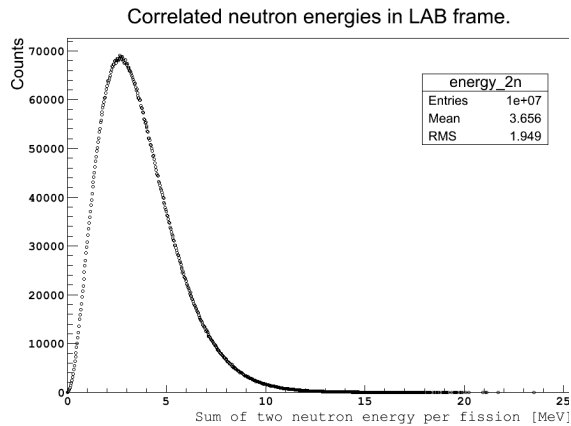

Figure 4.1: The energy distribution of sum of kinetic energy of two neutrons *a* and *b* emitted by fully accelerated fission fragments as seen in laboratory frame

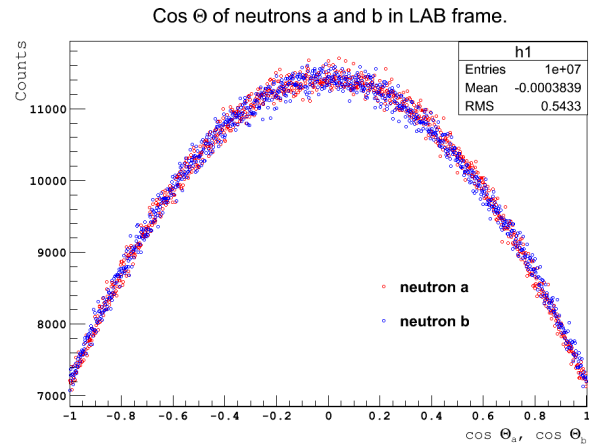

Figure 4.2: Angular distribution of prompt neutrons a (red) and b (blue) emitted by two fission fragments as seen in laboratory frame.

Angular distributions of prompt neutrons, as seen in laboratory frame, are presented in Fig. 4.2. That is, in principal, what everyone should expect for detection of one neutron. Here the neutron *a* is coming from one fission fragment and the neutron *b* is coming from the other one as was assumed above. First we note that angular distributions of both neutrons a and b look statistically similar as we can expect because there is no reason for a discrepancy. Also, as we can see, the resulting angular distribution is strongly anisotropic: more neutrons are emitted perpendicular to the beam directions (cos Θ = 0) than those in parallel (cos Θ = ±1). We can conclude here that the angular distribution of the fission fragments is strongly manifested in the angular distribution of prompt neutrons in laboratory frame. That result is important and could be used widely.

After energy and angular distributions of both neutrons *a* and *b* in the LAB frame were simulated, and we confirmed that our simulation is sensible, the next step is to investigate the two neutron correlation yield as a function of different quantities. We can count, for example, how many of them are going in anti-parallel directions and how many are going in parallel directions with respect to each other. Then the asymmetry can be calculated as was discussed earlier (formula 1.6). The results of two neutron correlation as a function of the

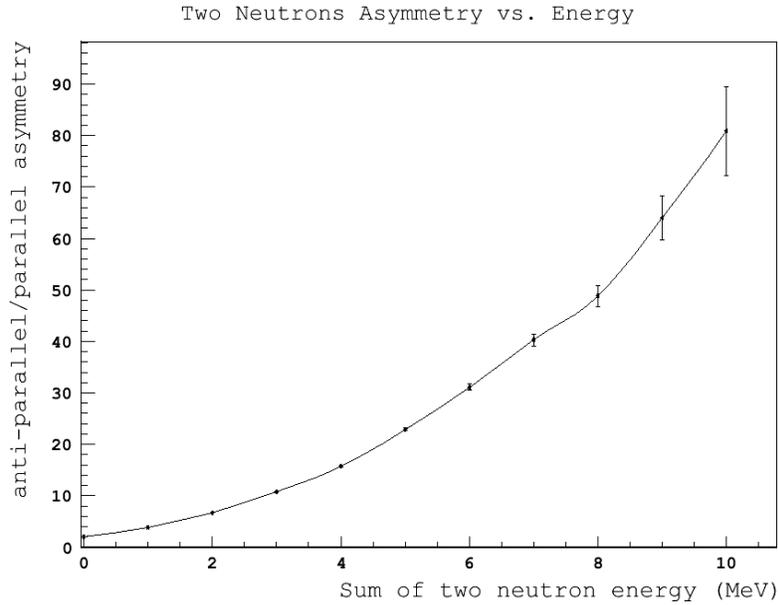

Figure 4.3: Calculated 2n asymmetry (antiparallel/parallel) as a function of the sum of two neutron energies

sum of two neutron energies are represented in Fig. 4.3. Here the following was assumed:

- two neutrons are antiparallel to each other if $\cos(\Theta_{2n}) < -0.9$
- two neutrons are parallel to each other if $\cos(\Theta_{2n}) > 0.9$

where $\Theta_{2n}$ is the calculated angle between neutrons a and b as seen in laboratory frame.

Of course, the events are binned in some energy interval. The exact values of intervals were used, numerical values of calculated 2n asymmetry and some other important quantities are shown in table 4.1. Here:

1. the "interval statistics" is the ratio of 2n pairs in some energy interval to the total number of 2n pairs.

2. the "asymmetry statistics" is the ratio of the total number of 2n pairs going in antiparallel and parallel directions in some energy interval to the total number of 2n pairs. That quantity is very useful and will be discussed widely. That is, in principal, the statistics we should expect for the total number of 2n coincidences (anti-parallel + parallel) assuming $4\pi$ detector geometry and the 100% absolute detector efficiency.

3. the "2n asymmetry" is the ratio of 2n pairs going in anti-parallel directions to 2n pairs going in parallel.

As we can see the resulting two neutron asymmetry is a strong function of the sum of two neutron energies. It increases from about 2 up to about 80 as we move from 0 to 10 MeV energy point. Also note, that at the higher energies the errors become significant and reach up to 10% value at the 10 MeV energy point. That is simply because the number of counts becomes smaller at the higher energy range. As we can see from table 4.1, at 10 MeV energy point, for example, the numbers of neutron pairs going in antiparallel and in parallel direction are about 7200 and 90 correspondingly. The asymmetry statistics at this point is about 0.07%. To study the asymmetries at the high energy interval, to reduce the error bars, high statistics are needed. The maximum asymmetry statistics of about 2.7% is reached at the (2 - 3) MeV energy interval and the corresponding asymmetry is about 10 here. It could be noted that asymmetry statistics qualitatively follow the energy spectrum of the sum of two neutron energies presented earlier in Fig. 4.1: it starts from about 0.4% at 0 MeV point, reaches the maximum value of about 2.7% at the energy range of (2 - 3) MeV, and goes down to 0.07% at 10 MeV point.

| #  | ($E_a+E_b$) MeV | 2n total | 2n interval | 2n antiparall | 2n parall. | interval statistics | asymmetry statistics | 2n asymmetry |
|----|-----------------|----------|-------------|---------------|------------|---------------------|----------------------|--------------|
| 0  | 0 - 1           | 10M      | 399953      | 27920         | 13556      | 0.0400              | 0.0041               | 2.06 ±0.02   |
| 1  | 1 - 2           | 10M      | 1628053     | 142211        | 36695      | 0.1628              | 0.0179               | 3.88 ±0.02   |
| 2  | 2 - 3           | 10M      | 2238223     | 232958        | 34691      | 0.2238              | 0.0268               | 6.72 ±0.04   |
| 3  | 3 - 4           | 10M      | 2048413     | 243893        | 22714      | 0.2048              | 0.0267               | 10.74± 0.07  |
| 4  | 4 - 5           | 10M      | 1514708     | 200007        | 12707      | 0.1515              | 0.0213               | 15.74± 0.14  |
| 5  | 5 - 6           | 10M      | 976912      | 140562        | 6133       | 0.0977              | 0.0147               | 22.92± 0.30  |
| 6  | 6 - 7           | 10M      | 571885      | 88873         | 2854       | 0.0572              | 0.0092               | 31.14± 0.59  |
| 7  | 7 - 8           | 10M      | 312163      | 51793         | 1287       | 0.0312              | 0.0053               | 40.24± 1.14  |
| 8  | 8 - 9           | 10M      | 160819      | 28005         | 574        | 0.0161              | 0.0029               | 48.79± 2.06  |
| 9  | 9 - 10          | 10M      | 79827       | 14595         | 228        | 0.0080              | 0.0015               | 64.01± 4.27  |
| 10 | 10 - 11         | 10M      | 37923       | 7201          | 89         | 0.0038              | 0.0007               | 80.91± 8.63  |

Table 4.1: Calculated 2n asymmetry (anti-parallel/parallel) as a function of the sum of two neutron energies

Also it would be interesting to calculate the two neutron asymmetry as a function of the energy cut on each neutron's energy and compare with the previous results. Such a kind of calculation is presented in Fig. 4.4 and in table 4.2.

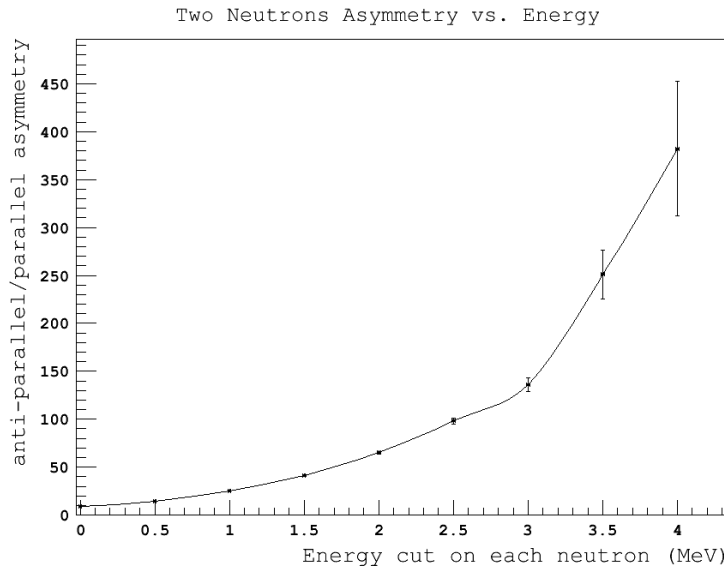

Figure 4.4: Calculated 2n asymmetry (antiparallel/parallel) as a function of the energy cut on each neutron energy

As we can see, by doing, say, 1 MeV energy cut on each neutron energy, the expected asymmetry would be about 25, and by doing 4 MeV energy cut the expected asymmetry reaches the high value of about 380. Nevertheless, the problem here is as before: the number of counts at high energy interval becomes smaller, which will increase the error bars in expected asymmetry values. It is interesting to note that even without any energy cut, by counting all neutron pairs going in antiparallel and in parallel directions, the expected asymmetry would be about 9, which corresponds to the maximum 2n asymmetry statistics of about 13%.

| # | $E_{cut}$ MeV | 2n total | 2n cut | 2n antiparall | 2n parall. | cut statistics | asymmetry statistics | 2n asymmetry |
|---|---|---|---|---|---|---|---|---|
| 0 | 0.0 | 10M | 10000000 | 1184431 | 131572 | 1.0000 | 0.1316 | 9.00 ± 0.03 |
| 1 | 0.5 | 10M | 7356078 | 962173 | 65184 | 0.7356 | 0.1027 | 14.76 ± 0.06 |
| 2 | 1.0 | 10M | 4429529 | 642068 | 25642 | 0.4430 | 0.0668 | 25.04 ± 0.16 |
| 3 | 1.5 | 10M | 2413642 | 382418 | 9285 | 0.2414 | 0.0392 | 41.19 ± 0.43 |
| 4 | 2.0 | 10M | 1227505 | 209578 | 3219 | 0.1228 | 0.0213 | 65.11 ± 1.16 |
| 5 | 2.5 | 10M | 592912 | 108071 | 1100 | 0.0593 | 0.0109 | 98.25 ± 2.98 |
| 6 | 3.0 | 10M | 275153 | 52865 | 388 | 0.0275 | 0.0053 | 136.25 ± 6.94 |
| 7 | 3.5 | 10M | 123314 | 25105 | 100 | 0.0123 | 0.0025 | 251.05 ± 25.15 |
| 8 | 4.0 | 10M | 53842 | 11469 | 30 | 0.0054 | 0.0011 | 382.30 ± 69.89 |

Table 4.2: Calculated 2n asymmetry (anti-parallel/parallel) as a function of the energy cut on each neutron energy

In general, the results presented in table 4.2 are better than results presented in table 4.1 in the following sense: in the first case (table 4.1), the maximum asymmetry statistics we can get is about 2.7%, which corresponds to the 2n asymmetry value of about 10. In the second case, by doing, say, 0.5 MeV energy cut on each neutron energy, we can easily reach the asymmetry statistics of about 10%, which corresponds to the 2n asymmetry value of about 15. By doing the energy cut we can significantly increase the calculated 2n asymmetry still having a small statistical error.

There are several ways to make the results discussed above more realistic. Some of them are directly following from the assumptions that were made in simulation so far:

- we can use the more realistic FF's mass distribution (Fig. 1.3) instead of uniform one.

- we can use the more realistic multiplicity value instead of assuming that each fission fragment emits just one neutron.

- we can estimate the neutron multiple scattering effect inside the target. That, of course, will decrease the calculated 2n asymmetry results.

That all can be done later, however, the results of simulation show the huge asymmetry effect in 2n correlation. That will potentially permit a new technique for actinide detection for homeland security and safeguards applications.

## 4.2 Count rate calculation

It was shown in the previous sections that the expected 2n asymmetry (antiparallel/parallel) is a high number (eqn. 1.7) and strongly depends on the sum of the two neutron energies (Fig. 4.3) as well as on the energy cut on each neutron (Fig. 4.4). For example, we can expect the asymmetry value of about 65 by doing a 2 MeV energy cut on each neutron still having reasonable asymmetry statistics of about 2%.

However, the problem arises as follows. Let us assume we have N fission events per beam pulse and let us count how many two neutron coincidences are true and how many are accidental ones. The true coincidences are between two neutrons coming from the same fission event and obviously for N fission events we will have N true coincidences for 100% efficiency. Let us call them $N_{true}$. The accidental coincidences are between two neutrons coming from the different fission events and, as can be easily seen, they are proportional to $N(N-1)$. Let us call them $N_{accidental}$. Now let us calculate the following ratio:

$$\frac{N_{true}}{N_{accidental} + N_{true}} = \frac{N}{N(N-1) + N} = \frac{1}{N} \tag{4.3}$$

To be able to observe the true coincidences we want the ratio above to be equal to one. The only way to do it is to make $N = 1$. That will guarantee that every coincidence will be a priori a true one with no way to have an accidental one. We need to design the experiment in such a way that the following condition is satisfied:

$$N = \frac{1 \text{ fission event}}{\text{pulse}} \tag{4.4}$$

Let us do the count rate calculation to check the possibility to satisfy the condition above. By taking $\tau = 20 \text{ } ns$ pulse width and $I = 20 \text{ } mA$ peak current, the number of electrons per pulse will be:

$$N_{e^-} = 20 \cdot 10^{-3} \frac{\text{Coloumb}}{\text{sec}} \times \frac{1 \text{ } e^-}{1.6 \cdot 10^{-19} \text{ Coloumb}} \times 20 \text{ ns} = 2.5 \cdot 10^9 \frac{e^-}{\text{pulse}} \tag{4.5}$$

To be specific, let us use the $^{235}U$ as a target. Fig. 4.5 shows $(\gamma, f)$ and $(\gamma, 2n)$ photo-nuclear cross sections as a function of incident photon energy [25]. As can be seen, the optimal energy of incident gammas would be about 6-7 MeV in the following sense: first, the $(\gamma, f)$ cross section is low in this energy interval, making it possible to satisfy the desired condition of having one impulse per pulse. Second, there is no way to have the direct "2n knockout" because we are well below the threshold energy of about 12 MeV for the $(\gamma, 2n)$ channel. By choosing, say, the 7 MeV electron beam energy, we will be able to study the pure $(\gamma, f)$ channel.

The bremsstrahlung spectrum with 7 MeV endpoint energy for the thin Al radiator is shown in Fig 4.6. That will produce about 0.05 photons/$e^-$/MeV/r.l. in the 6-7 MeV region.

Taking the thickness of Al radiator equal to 90 microns (about $10^{-3}$ radiation length), the number of bremsstrahlung photons going out of radiator in the 6-7 MeV energy range can be calculated as follow:

$$N_\gamma^I = 2.5 \cdot 10^9 \frac{e^-}{\text{pulse}} \times 0.05 \frac{\text{photons}}{e^- \text{ MeV r.l.}} \times 1 \text{ MeV} \times 10^{-3} \text{ r.l.} = 1.25 \cdot 10^5 \frac{\gamma's}{\text{pulse}} \tag{4.6}$$

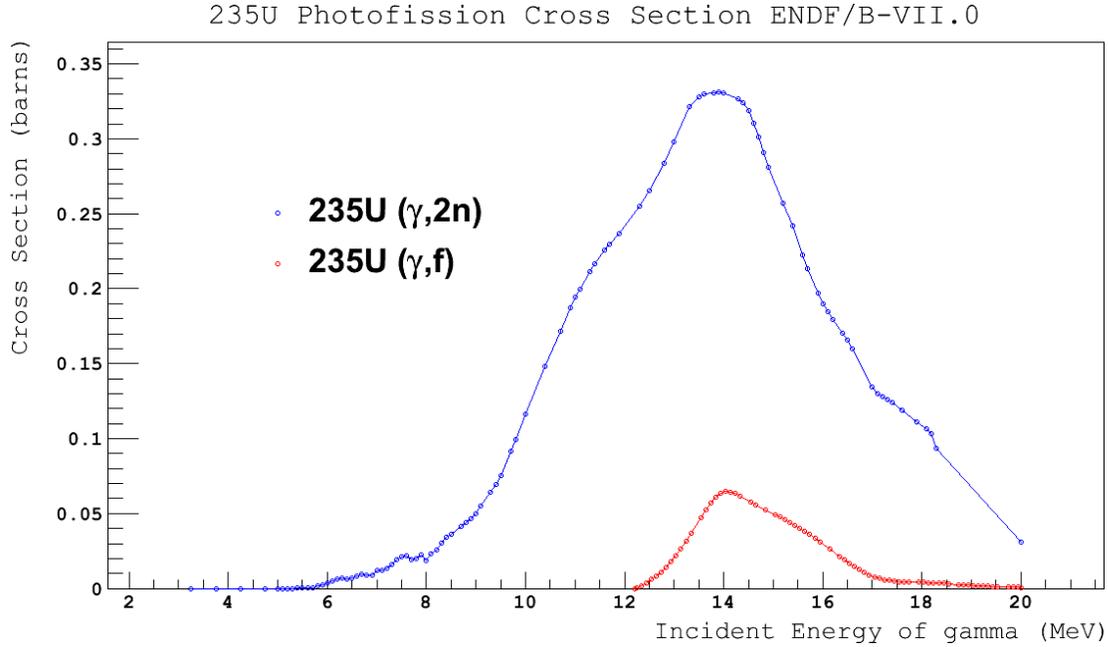

Figure 4.5: $^{235}U$ photofission cross section taken from ENDF/B-VII.0

Not all photons calculated above will hit the target. Some of them will be lost due to collimation. Assuming the collimation factor is about 50%, the number of photons hitting the target becomes:

$$N_Y = N_{Y'} \times 50\% = 6.25 \cdot 10^4 \; \frac{\gamma's}{pulse} \qquad (4.7)$$

We want one fission per pulse. That can be found by adjusting the target thickness from the equation below:

$$\frac{1 \; fission}{pulse} = N_Y \times t \times \sigma \qquad (4.8)$$

where $t$ the is the target thickness in atoms/cm$^2$ and the $\sigma$ is the ($\gamma, 2n$) photo-nuclear cross section and is about 7 mb/atom in the 6-7 MeV energy range as can be seen from Fig 4.5 above. The thickness becomes:

$$t \; \left[\frac{atoms}{cm^2}\right] = \frac{1 \frac{fission}{pulse}}{6.25 \cdot 10^4 \frac{\gamma's}{pulse} \times 7 \frac{mb}{atom}} = 2.29 \cdot 10^{21} \; \frac{atoms}{cm^2} \qquad (4.9)$$

and could be converted into cm as follows:

$$t \, [cm] = \frac{t \cdot M}{\rho \cdot N_A} = \frac{2.29 \cdot 10^{21} \frac{atoms}{cm^2} \times 235.04 \frac{g}{mol}}{19.1 \frac{g}{cm^3} \times 6.02 \cdot 10^{23} \frac{atoms}{mol}} = 470 \; \mu m \qquad (4.10)$$

where $M$ is the molar mass, $\rho$ is the density of $^{235}U$ and $N_A$ is the Avogadro number.

In the last step we were able, by varying the target thickness, to satisfy the desired situation of having the one fission per pulse. In principal, the other elements of the beam line, like the thickness of the radiator or the collimation hole, can be varied as well. After a reasonable judgment about the beam line elements is done, we still have the possibility to adjust the count rates by varying the LINAC beam parameters, such as the electron pulse width and the electron peak current.

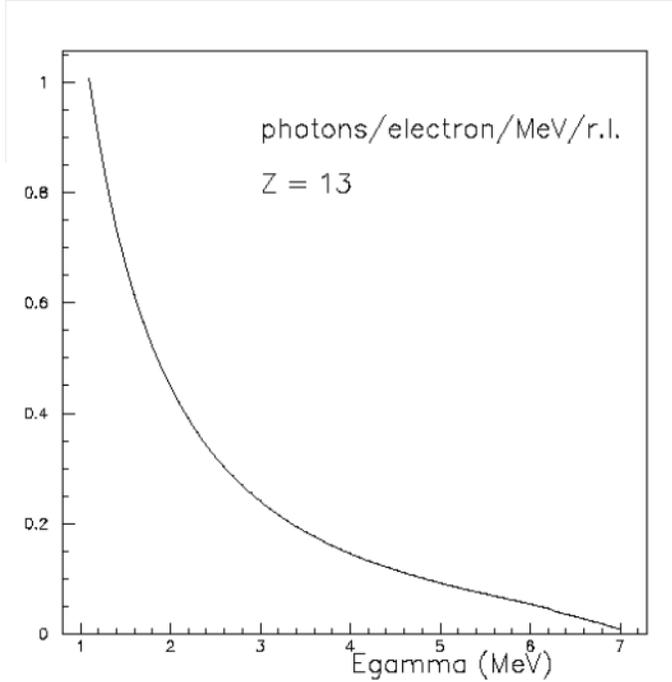

Figure 4.6: Bremsstrahlung spectrum of photons produced by 7 MeV electrons hitting the Al radiator

## 4.3 Beam time calculation

Let us now estimate the time needed to run the experiment. As was already mentioned in the previous section, to eliminate the accidental coincidences, the one fission per pulse condition is needed. Because of that, the High Repetition Rate Linac (HRRL) available at IAC will be a good choice.

The count rate for two neutron detectors, located 2 m away from the target as shown in Fig 4.7, can be estimated as follows:

$$N \left[\frac{\text{counts}}{\text{sec}}\right] = \frac{1 \text{ fission}}{\text{pulse}} \cdot N_G^2 \cdot N_{intr}^2 \cdot N_{cut} \cdot 2.2 \cdot 10^3 \text{ Hz} \quad (4.11)$$

where $N_G$ is the geometrical detector efficiency, $N_{intr}$ is the absolute intrinsic detector efficiency, $N_{cut}$ is the efficiency of the energy cut, 2.2 is the average number of neutrons per fission, $10^3$ Hz is the HRRL repetition rate.

The geometrical detector efficiency $N_G$ is proportional to the solid angle from which the target see the detector and can be calculated as follows:

$$N_G = \frac{\Omega}{4\pi} = \frac{S}{4\pi r^2} = \frac{(15 \times 88) \text{ cm}^2}{4\pi(2 \text{ m})^2} = \frac{0.132 \text{ m}^2}{50.258 \text{ m}^2} = 2.6 \cdot 10^{-3} \text{ sr} \quad (4.12)$$

.

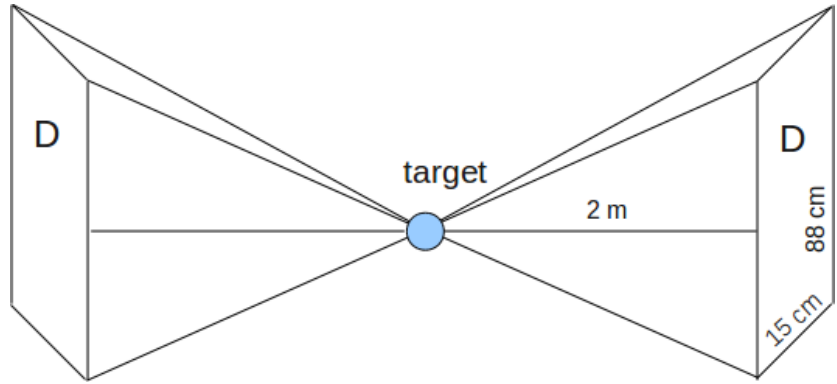

Figure 4.7: Two detector geometry located 2 m away from target

The intrinsic detector efficiency $N_{intr}$ can be conservatively assumed to be about 25%. The efficiency of energy cut $N_{cut}$ can be estimated from table 4.2 and, for 1 MeV energy cut, is about 44%. Substituting all the values above in formula 4.11, the count rate for two detectors becomes:

$$N = \frac{1\ \text{fission}}{\text{pulse}} \cdot (2.6 \cdot 10^{-3})^2 \cdot (0.25)^2 \cdot 0.44 \cdot 2.2 \cdot 10^3\ \text{Hz} = 4 \cdot 10^{-4}\ \frac{\text{counts}}{\text{sec}} \quad (4.13)$$

There are a total of 16 neutron detectors available for that experiment at the present time, which gives to us the factor of 8 × 8 = 64 and the total count rate becomes:

$$N_{16\ det} = N \times 64 = 2.6 \times 10^{-2}\ \frac{\text{counts}}{\text{sec}} \quad (4.14)$$

The expected statistics for one working day finally becomes:

$$N_{day} = N_{16\ det} \times 60\ \text{sec} \times 60\ \text{min} \times 8\ \text{hours} \approx 750\ \frac{\text{counts}}{\text{day}} \quad (4.15)$$

which is good enough for future analysis of the experimental data.

# Chapter 5

# Summary, conclusion

Below are the short summary and conclusion of the proposed two neutron correlation study in photofission of actinides:

- There is a need for in experimental data of two neutron correlation measurements in photofission of different materials.

- The preliminary calculation of the two neutron correlation shows a huge asymmetry effect: many more neutrons are emitted anti-parallel to each other than parallel to each other. That asymmetry becomes even more if the energy cut on each neutron is done. There are some factors, neutron multiple scattering effect, for example, that will reduce the calculated asymmetry and that could be calculated later. But that will not reduce the expected asymmetry significantly.

- We propose to measure and analyze the two neutron correlation yield resulting from two FF's as a function of different targets, the angle between the two neutrons and the neutron energies. There are a total of 16 "big" plastic detectors available at the present time, which can be used for neutron detection. With the High Repetition Rate Linac we can get about 750 counts per day.

- This study will potentially permit a new technique for actinide detection for homeland security and safeguards applications as well as improve our knowledge of correlated neutron emission.